\documentclass[reprint,superscriptaddress,showkeys,amsmath,amssymb,longbibliography,aps,pra,floatfix]{revtex4-1}

\usepackage{amsmath}
\usepackage{graphicx}
\usepackage{bm}
\usepackage{dcolumn}
\usepackage{hyperref}

\newcommand{\eps}{\varepsilon}

\newcommand{\kp}{\mathbf{k}\cdot\mathbf{p}}
\newcommand{\Egap}{E_{g}}
\newcommand{\Ep}{E_{P}}
\newcommand{\Esoc}{\Delta_{\text{soc}}}

\newcommand{\Hkp}{h_{\mathbf{k}\cdot\mathbf{p}}}

\newcommand{\Vconf}{V_{\text{ext}}}

\newcommand{\epseff}{\eps_{\text{eff}}}
\newcommand{\epsopt}{\eps_{\text{opt}}}

\newcommand{\epsin}{\eps_{\text{in}}}
\newcommand{\epsout}{\eps_{\text{out}}}

\newcommand{\Ket}[1]{ | #1 \rangle }

\newcommand{\BraOperKet}[3]{
\langle #1 | #2 | #3 \rangle
}

\newcommand{\Sixj}[6]{
\left\{ \begin{matrix} #1 & #2 & #3 \\ #4 & #5 & #6 \end{matrix} \right\}
}
\newcommand{\Threej}[6]{
\left( \begin{matrix} #1 & #2 & #3 \\ #4 & #5 & #6 \end{matrix} \right)
}
\newcommand{\RME}[3]{
\langle #1 \Vert \, #2 \Vert #3 \rangle
}

\begin{document}

\title{One-photon absorption by inorganic perovskite nanocrystals: A theoretical
study}

\author{T. P. T. Nguyen}
\email{phuctan3108@gmail.com}

\author{S. A. Blundell}
\email{steven.blundell@cea.fr}

\affiliation{Univ.\ Grenoble Alpes, CEA, CNRS, IRIG, SyMMES, F-38000 Grenoble,
France}

\author{C. Guet}
\email{cguet@ntu.edu.sg}

\affiliation{Energy Research Institute, Nanyang Technological University, 637141
Singapore}
\affiliation{School of Materials Science and Engineering, Nanyang Technological
University, 639798 Singapore}

\date{\today}

\begin{abstract}
The one-photon absorption cross section of nanocrystals (NCs) of the
inorganic perovskite CsPbBr$_{3}$ is studied theoretically using
a multiband $\mathbf{k}\cdot\mathbf{p}$ envelope-function model combined
with a treatment of intercarrier correlation by many-body perturbation
theory. A confined exciton is described first within the Hartree-Fock
(HF) approximation, and correlation between the electron and hole
is then included in leading order by computing the first-order vertex
correction to the electron-photon interaction. The vertex correction
is found to give an enhancement of the near-threshold absorption cross
section by a factor of up to 4 relative to the HF (mean-field) value
of the cross section, for NCs with an edge length $L=9$--12~nm
(regime of intermediate confinement). The vertex-correction enhancement
factors are found to decrease with increasing exciton energy; the
absorption cross section for photons of energy $\omega=3.1$~eV (about
0.7~eV above threshold) is enhanced by a factor of only 1.4--1.5
relative to the HF value. The $\kp$ corrections to the absorption
cross section are also significant; they are found to increase the
cross section at an energy $\omega=3.1$~eV by about 30\% relative
to the value found in the effective-mass approximation. The theoretical
absorption cross section at $\omega=3.1$~eV, assuming a Kane parameter
$E_{P}=20$~eV, is found to be intermediate among the set of measured
values (which vary among themselves by nearly an order of magnitude)
and to obey a power-law dependence $\sigma^{(1)}(\omega)\propto L^{2.9}$
on the NC edge length $L$, in good agreement with experiment. The
dominant contribution to the theoretical exponent 2.9 is shown to
be the density of final-state excitons. We also calculate the radiative
lifetimes of the ground-state $1S_{e}$-$1S_{h}$ exciton of NCs of
CsPbBr$_{3}$ and CsPbI$_{3}$, finding an overestimate by a factor
of up to about two (for $\Ep=20$~eV and 17~eV, respectively) compared
to the available experimental data, which vary among themselves by
about $\pm40$\%. The sources of theoretical uncertainty and the possible
reasons for the discrepancies with experiment are discussed. The main
theoretical uncertainty in these calculations is in the value of the
Kane parameter $E_{P}$.
\end{abstract}


\maketitle

\section{\label{sec:introduction} Introduction}

In 2015, Protesescu \emph{et al}.\ \cite{ProtesescuNanoLett2015}
reported a novel class of semiconductor nanocrystal (NC) materials
with outstanding emission and absorption properties. These were NCs
of all-inorganic lead halide perovskites CsPbX$_{3}$ (X = Cl, Br,
I). The NCs fluoresce strongly, with quantum yields approaching 100\%
\cite{krieg-18-sqd}, and the emission frequency is tunable over the
whole visible spectrum by varying the size and halide composition
X (including mixtures of different halides) \cite{ProtesescuNanoLett2015}.
The emission rate is one to two orders of magnitude faster than any
other known semiconductor NC at room temperature, and about three
orders of magnitude faster at cryogenic temperatures \cite{raino-16-sqd,BeckerNatLett2018}.
Important recent applications of these NCs have been made to lasers
\cite{pan-15-sqd,yakunin-15-sqd}, light-emitting diodes \cite{deng-16-sqd,li-16-sqd},
and room-temperature single-photon sources \cite{utzat-19-sqd}.

The fast emission of NCs of CsPbX$_{3}$ is thought to be related
to the existence of a bright triplet ground-state exciton in these
materials, in contrast to the dark (poorly emitting) ground-state
exciton found in all other known inorganic semiconductor NCs \cite{BeckerNatLett2018}.
This would explain, for instance, the persistence of the bright emission
down to cryogenic temperatures \cite{raino-16-sqd}. It has been speculated
that the existence of the bright ground state could be related to
a strong Rashba spin-orbit coupling in the NCs \cite{BeckerNatLett2018,SercelNanoLett2019},
which can lead to an inversion of the usual ground-state exciton fine-structure
energy ordering, with the dark-exciton fine-structure state above
the bright state. These issues have stimulated much recent theoretical
work on the exciton fine structure \cite{BeckerNatLett2018,ben-aich-19-sqd,SercelNanoLett2019,sercel-19a-psk}
and on the ground-state radiative decay rates \cite{BeckerNatLett2018}
of NCs of CsPbX$_{3}$.

Absorption by NCs of CsPbX$_{3}$ has also been extensively studied
experimentally. One-photon \cite{WangAdvMater2015,makarov-16-sqd,xu-16-sqd,chen-17a-sqd,nagamine-18-sqd},
two-photon \cite{chen-17b-sqd,chen-17a-sqd,nagamine-18-sqd,pramanik-19-sqd},
and up to five-photon \cite{chen-17b-sqd} absorption cross sections
have recently been measured. Less attention has been given theoretically
to absorption by these NCs, however. In this paper, we calculate the
one-photon absorption cross section for NCs of CsPbBr$_{3}$ and compare
with the available measurements.

The paper is organized as follows. In Sec.~\ref{sec:formalism} we
outline our multiband $\kp$ envelope-function formalism. As we will
see, $\kp$ corrections to the absorption cross section are surprisingly
large. Therefore, our approach is based on a $4\times4$ $\kp$ model,
containing the highest-lying valence band (VB) and the lowest-lying
conduction band (CB). We discuss this model in Sec.~\ref{subsec:model}.
Also important for emission and absorption in NCs of CsPbX$_{3}$
are the large intercarrier correlation corrections that are found,
especially for the ground-state exciton. We treat correlation using
methods of many-body perturbation theory (MBPT). This involves starting
in lowest order with a self-consistent Hartree-Fock (HF) model and
then applying Coulomb correlation corrections. This formalism is discussed
in Secs.~\ref{subsec:e-ph-mxel} and \ref{subsec:vertex-correction}.
A important feature of our numerical approach is the use of a spherical
basis set (applying to a spherically symmetric confining potential)
to accelerate the calculation of the correlation corrections. In Appendix~\ref{app:momentum-mxel},
we derive a key formula, used extensively in the calculations, for
the reduced momentum matrix element in the $4\times4$ $\kp$ model
for states of spherical form.

In Sec.~\ref{sec:results} we then apply these methods to emission
and absorption in inorganic perovskite NCs. A difficulty with these
materials, which have only recently become the subject of intensive
research, is that many of the material parameters are at present uncertain.
This includes effective masses and the Kane parameter, the latter
controlling the strength of the electron-photon coupling for interband
transitions. Hence, in Sec.~\ref{subsec:parameters}, we first discuss
the available data and our choice of parameters. Although the main
focus of the paper is absorption, there are important related data
on the radiative lifetimes of the ground-state bright excitons. Therefore,
in Sec.~\ref{subsec:results-lifetimes} we first apply our methods
to calculate radiative lifetimes. The calculations of one-photon absorption
then follow in Sec.~\ref{subsec:absorption-spectra}. Our conclusions
are given in Sec.~\ref{sec:conclusions}.

We use atomic units throughout in all formulas.

\section{\label{sec:formalism} Formalism}

\subsection{\label{subsec:model}Model}

We use a multiband envelope-function formalism for a system of carriers
(electrons and holes) confined by a mesoscopic potential $V_{\text{ext}}$
\cite{Kira&Koch}. The bulk band structure is given by a $\kp$ Hamiltonian
$\Hkp$ and the Coulomb interactions among the carriers are screened
by the dielectric constant $\varepsilon_{\text{in}}$ of the NC material.
The system Hamiltonian (in the space of electron envelope functions)
is then
\begin{eqnarray}
H & = & \sum_{ij}\{i^{\dagger}j\}\BraOperKet{i}{\Hkp+\Vconf}{j}\nonumber \\
 &  & {}+\frac{1}{2}\sum_{ijkl}\{i^{\dagger}j^{\dagger}lk\}\BraOperKet{ij}{g_{12}}{kl}\,,\label{eq:hamiltonian}
\end{eqnarray}
where the notation $\{i_{1}^{\dagger}i_{2}^{\dagger}\ldots j_{1}j_{2}\ldots\}$
indicates a normally ordered product of creation (and absorption)
operators for electron envelope states $i_{1},i_{2}\ldots$ (and $j_{1},j_{2}\ldots$),
and the sums span all states in all bands (conduction or valence)
included in the calculation. We include only the long-range (LR) Coulomb
interaction in this work,
\begin{eqnarray}
g_{12} & = & \frac{1}{\epsin|\mathbf{r}_{1}-\mathbf{r}_{2}|}\,,\label{eq:lrcoul}
\end{eqnarray}
where $\epsin$ is the dielectric constant of the NC material appropriate
to the length scale $L_{\text{dot}}$ of the nanostructure (see Sec.~\ref{subsec:parameters}).
For the applications of this paper, the short-range (SR) Coulomb interaction
\cite{Knox,[{}] [{ [Sov.\ Phys.\ JETP \textbf{33}, 108 (1971)].}] PikusJETP1971}
is suppressed relative to the LR term by a factor of order $(L_{\text{atom}}/L_{\text{dot}})^{2}$,
where $L_{\text{atom}}$ is the atomic length scale, and can be neglected.
The LR Coulomb interaction~(\ref{eq:lrcoul}) is in principle modified
by the mismatch with the dielectric constant $\epsout$ of the surrounding
medium \cite{karpulevich-19}, which leads to induced polarization
charges at the interface \cite{Jackson}, although we will not consider
this effect in the present paper.

NCs of inorganic perovskite are generally cuboid \cite{ProtesescuNanoLett2015}.
Nevertheless, for reasons of computational efficiency, we will choose
the basis states $i$, $j$, \ldots , etc., appearing in Eq.~(\ref{eq:hamiltonian})
to be those for a spherical confining potential $\Vconf^{\text{sph}}$.
This choice is particularly advantageous in many-body calculations.
The integrals over angles in matrix elements can be handled analytically,
so that only radial integrals remain to be evaluated numerically.
Moreover, in the sums over virtual states appearing in MBPT, it is
possible to sum over the magnetic substates analytically \cite{LindgrenMorrison,Brink&Satchler},
thereby reducing substantially the effective size of the basis set.
The nonspherical part of the confining potential $V_{\text{ext}}^{\text{ns}}=V_{\text{ext}}-V_{\text{ext}}^{\text{sph}}$
(which can include terms arising from the crystal lattice as well
as from the overall shape of the NC) can in principle be treated as
a perturbation in later stages of the calculation procedure.

To generate a spherical basis, we take $\Vconf^{\text{sph}}$ to be
a spherical well with infinite walls, 
\begin{equation}
\Vconf^{\text{sph}}(r)=\left\{ \begin{matrix}0\text{, if }r<R\\
\infty\text{, otherwise}
\end{matrix}\right.\,.\label{eq:sphericalWell}
\end{equation}
If the NC is a cube with edge length $L$, we choose the radius $R$
to satisfy 
\begin{equation}
R=L/\sqrt{3}\,.\label{eq:radiusL}
\end{equation}
The above choice of $R$ ensures that the ground-state eigenvalue
for noninteracting electrons in the cube and the sphere is the same
\cite{sercel-19a-psk,nguyen-20-sqd}. In fact, as discussed in Ref.~\cite{nguyen-20-sqd},
the energies of noninteracting excited $nS$ and $nP$ states in the
sphere of radius $R$ also agree closely, to within a few percent,
with the energies of the analogous `$S$-like' and `$P$-like' states
\cite{shaw-74-sqd} in the cube of edge length $L$.

Matrix elements can also be reproduced quite accurately using the
radius~(\ref{eq:radiusL}). In Ref.~\cite{nguyen-20-sqd}, it is
shown that the first-order Coulomb energy for the ground-state exciton
differs between the sphere and the cube by only about 1.5\%. Moreover,
the interband matrix element for the radiative decay of a single exciton
depends on a simple overlap of the electron and hole envelope functions
\cite{efros-82-sqo,Kira&Koch} (see also Appendix~\ref{app:momentum-mxel}).
Since the ground-state electron and hole wave functions are approximately
equal, this overlap is close to unity, independently of whether the
confining potential is a cube or a sphere. For these reasons, in this
work we shall make the approximation of neglecting the cubic correction
terms in $V_{\text{ext}}^{\text{ns}}$ entirely; as we will see, there
are other theoretical uncertainties that are at present likely to
be larger.

We use a $4\times4$ $\kp$ model, which includes the $s$-like VB
($R_{6}^{+}$) and the $p_{1/2}$-like CB ($R_{6}^{-}$) around the
$R$ point of the Brillouin zone of the inorganic perovskite compounds
\cite{Efros&RosenPRB1998,even-14b-sqd,BeckerNatLett2018}. For spherical
confinement, the angular part of an envelope function with orbital
angular momentum $l$ couples to a Bloch function with Bloch angular
momentum $J$ (here $J=1/2$) to give a state with total angular momentum
$(F,m_{F})$ \cite{ekimov-93-sqd}. We will denote this state by a
basis vector $\Ket{(l,J)Fm_{F}}$. In the $4\times4$ $\kp$ model,
the total wave function (including envelope and Bloch functions) can
then be written as a sum of two components, 
\begin{equation}
\Ket{\eta Fm_{F}}=\frac{g_{s}(r)}{r}\Ket{(l,1/2)Fm_{F}}+\frac{\bar{g}_{p}(r)}{r}\Ket{(\bar{l},1/2)Fm_{F}}\,\label{eq:4compState}
\end{equation}
{[}see Ref.~\cite{ekimov-93-sqd}, with the terms for the $p_{3/2}$-like
($R_{8}^{-}$) band dropped{]}. Here $g_{s}(r)$ and $\bar{g}_{p}(r)$
are the radial envelope functions for the $s$-like and $p_{1/2}$-like
bands, respectively. The allowed values of the angular momenta $l$
and $\bar{l}$ follow from angular-momentum and parity selection rules
\cite{ekimov-93-sqd}.

For states in the $p_{1/2}$-like CB, the term involving $\bar{g}_{p}(r)$
in Eq.~(\ref{eq:4compState}) is the large component of the wave
function, while the other term is a small component representing the
admixture of the VB state into the CB state due to the finite range
of the confining potential $\Vconf$ and the $\kp$ interaction. In
VB states, the small and large components are reversed. Because of
the small components, the formalism picks up the leading $\kp$ corrections
arising from the coupling of the CB and VB. We will conventionally
label spherical states~(\ref{eq:4compState}) by the quantum numbers
of the large component. For instance, an electronic (CB) $(1P_{3/2})_{e}$
state has $\bar{l}=1$, $l=2$, and $F=3/2$, corresponding to a small
component with $D_{3/2}$ symmetry, while a hole (VB) $1S_{h}$ state
has $l=0$, $\bar{l}=1$, and $F=1/2$.

The first step in the many-body procedure is to solve the self-consistent
HF equations including exact exchange \cite{LindgrenMorrison,ShavittBartlett}
for the spherical $4\times4$ $\kp$ model \cite{nguyen-20-sqd}.
The single-particle basis states of the many-body procedure (Sec.~\ref{subsec:vertex-correction})
are then calculated in this HF potential. Specifically, we first solve
the HF equations for the $1S_{e}$-$1S_{h}$ ground-state exciton.
The dominant term for this system is the direct Coulomb interaction
between the electron and hole; the exchange term, although included,
is a small correction term formally of order $(\kp)^{2}$ or $(L_{\text{atom}}/L_{\text{dot}})^{2}$.
Next, a set of excited (unoccupied) HF states is generated up to a
high energy cutoff. Together with the occupied $1S_{e}$ and $1S_{h}$
states, this set forms a complete HF basis for MBPT. The HF potential
here is defined as in Ref.~\cite{nguyen-20-sqd} via a configuration
average. With this definition, the HF potential for the excited electron
states is due entirely to the $1S_{h}$ state, while that for the
excited hole states is due entirely to the $1S_{e}$ state. In this
way, an approximation to the electron-hole Coulomb energy is built
into the eigenvalues of the basis set.

\subsection{\label{subsec:e-ph-mxel}Lifetime and absorption cross section}

Expressions for the one-photon emission and absorption rates for NCs
are readily found using time-dependent perturbation theory (see, for
example, Refs.~\cite{elliott-57-sqd,efros-82-sqo,hu-90-sqd}). The
radiative decay rate for a general single-exciton state $(e,h)$ (with
total angular momentum $F_{\text{tot}}=1$) can be written
\begin{equation}
\frac{1}{\tau_{eh}}=\frac{4}{9}\frac{n_{\text{out}}\omega_{eh}}{c^{3}}f_{\varepsilon}^{2}\left|M_{eh}\right|^{2}\,.\label{eq:decay-rate}
\end{equation}
Here, $\omega_{eh}$ is the energy of emitted photon (the exciton
energy), $n_{\text{out}}=\sqrt{\varepsilon_{\text{out}}}$ is the
refractive index of the medium surrounding the NC, with $\varepsilon_{\text{out}}$
the dielectric constant of this medium, $f_{\varepsilon}$ is the
dielectric screening factor (discussed further below), and $M_{eh}$
is the reduced amplitude for the decay \footnote{We write all reduced amplitudes as absorption amplitudes, which are
the same as the corresponding emission amplitudes up to a phase factor.
The radiative decay rate is unaffected, since it depends on the modulus
squared $|M_{eh}|^{2}$.},
\begin{equation}
M_{eh}=\RME{(e,h)F_{\text{tot}}}{P^{1}}{\Psi_{0}}\,.\label{eq:red-amplitude}
\end{equation}
The state $\Ket{(e,h)F_{\text{tot}}}$ here is the (correlated) exciton
state, $\Ket{\Psi_{0}}$ is the ground state of the NC (also correlated),
and $P^{1}$ is the total momentum operator.

The one-photon absorption cross section for frequency $\omega$ (for
absorption from the ground state to a single exciton) is given by
\cite{elliott-57-sqd,efros-82-sqo,hu-90-sqd}
\begin{equation}
\sigma^{(1)}(\omega)=\frac{4\pi^{2}}{3}\frac{f_{\varepsilon}^{2}}{n_{\text{out}}c\omega}\sum_{eh}\left|M_{eh}\right|^{2}\Delta_{eh}(\omega-\omega_{eh})\,.\label{eq:cross-section}
\end{equation}
The total cross section contains a sum over all possible exciton final
states $(e,h)$, with each transition broadened by a normalized line-shape
function $\Delta_{eh}(\omega-\omega_{eh})$ satisfying 
\begin{equation}
\int_{0}^{\infty}\!\Delta_{eh}(\omega-\omega_{eh})\,d\omega=1\,,\label{eq:g-norm}
\end{equation}
which is discussed further in Sec.~\ref{subsec:absorption-spectra}.

The factor $f_{\varepsilon}$ in Eqs.~(\ref{eq:decay-rate}) and
(\ref{eq:cross-section}) relates the photon electric field inside
the NC to the photon electric field at infinity. For a spherical NC,
the electric field $\mathbf{E}_{\text{in}}$ inside is parallel to
the electric field $\mathbf{E}_{\infty}$ at infinity and has a constant
value, independent of the position inside the NC (in the electrostatic
approximation) \cite{Jackson}. The dielectric screening factor is
then defined as $f_{\varepsilon}=|\mathbf{E}_{\text{in}}|/|\mathbf{E}_{\infty}|$,
which has the value for a sphere \cite{Jackson} 
\begin{equation}
f_{\varepsilon}^{\text{sph}}=\frac{3\varepsilon_{\text{out}}}{\varepsilon_{\text{in}}'+2\varepsilon_{\text{out}}}\,,\label{eq:feps-sphere}
\end{equation}
where $\varepsilon_{\text{in}}'$ is the \emph{optical} dielectric
constant of the NC material, which can have a different value from
the dielectric constant $\varepsilon_{\text{in}}$ used to screen
the Coulomb interactions in Eq.~(\ref{eq:lrcoul}). The case of a
cubic NC, which is found for inorganic perovskites, can be handled
numerically \cite{BeckerNatLett2018}. For a cube, the internal electric
field $\mathbf{E}_{\text{in}}$ is not in general parallel to $\mathbf{E}_{\infty}$,
and its magnitude varies over the volume of the NC. However, we shall
here use a similar formalism as for a sphere and define $f_{\varepsilon}$
to be a suitable constant average value for $|\mathbf{E}_{\text{in}}|/|\mathbf{E}_{\infty}|$.
Thus, $f_{\varepsilon}$ can be removed from the integral over electron
coordinates in the matrix element $M_{eh}$ (where, more generally,
it should appear \cite{BeckerNatLett2018}), as we have done in Eqs.~(\ref{eq:decay-rate})
and (\ref{eq:cross-section}). According to the numerical calculations
in Ref.~\cite{BeckerNatLett2018}, the average value of $f_{\varepsilon}$
for a cube is about 6\% smaller than $f_{\varepsilon}^{\text{sph}}$
for the parameters used here (see Sec.~\ref{subsec:results-lifetimes}).

An expression for the reduced amplitude~(\ref{eq:red-amplitude})
in lowest order (at HF level) can be obtained as follows. The lowest-order
exciton state can be written 
\begin{multline}
\Ket{(e,h)F_{\text{tot}}m_{\text{tot}}}=\sum_{m_{e}m_{h}}(-1)^{F_{h}-m_{h}}\\
\times\left\langle F_{e}m_{e},F_{h}{-m_{h}}\kern0.1em \right|\!\left.F_{\text{tot}}m_{\text{tot}}\right\rangle \{e_{m_{e}}^{\dagger}h_{m_{h}}\}\Ket{0}\,,\label{eq:Xeh0}
\end{multline}
where $\Ket{0}$ is the effective vacuum (no carriers present), and
$\left\langle F_{e}m_{e},F_{h}{-m_{h}}\kern0.1em \right|\!\left.F_{\text{tot}}m_{\text{tot}}\right\rangle $
is a Clebsch-Gordon coefficient for coupling the electron and hole
angular momenta to a total angular momentum $F_{\text{tot}}$. (The
minus sign in $-m_{h}$ and the phase factor are necessary because
the hole is associated with an absorption operator \cite{Edmonds}.)
The effective vacuum $\Ket{0}$ is also the lowest-order approximation
to $\Ket{\Psi_{0}}$ (which in principle can contain correlation corrections
from virtual excitons). Substituting Eq.~(\ref{eq:Xeh0}) into Eq.~(\ref{eq:red-amplitude}),
one finds the lowest-order reduced amplitude 
\begin{equation}
M_{eh}^{(0)}=\delta(F_{\text{tot}},1)\RME{F_{e}}{p^{1}}{F_{h}}\,,\label{eq:M0-red}
\end{equation}
where $\RME{F_{e}}{p^{1}}{F_{h}}$ is a single-particle reduced momentum
matrix element. In Appendix~\ref{app:momentum-mxel}, we derive an
expression for a general reduced matrix element $\RME{F_{a}}{p^{1}}{F_{b}}$
in the $4\times4$ $\kp$ model with states $a$ and $b$ of spherical
form~(\ref{eq:4compState}). This expression has the form of radial
integrals and angular factors, and includes all $\kp$ corrections
arising from the small components of the single-particle states.

The factor $\delta(F_{\text{tot}},1)$ in Eq.~(\ref{eq:M0-red})
embodies the basic selection rule for one-photon recombination that
the initial state must have $F_{\text{tot}}=1$ to conserve angular
momentum. Further selection rules are associated with the reduced
matrix element $\RME{F_{e}}{p^{1}}{F_{h}}$ (see Appendix~\ref{app:momentum-mxel}).

In the next section, we consider the first-order corrections to $M_{eh}^{(0)}$
arising from Coulomb correlation.

\subsection{\label{subsec:vertex-correction}First-order Coulomb correlation}

\begin{figure}
\includegraphics[scale=0.33]{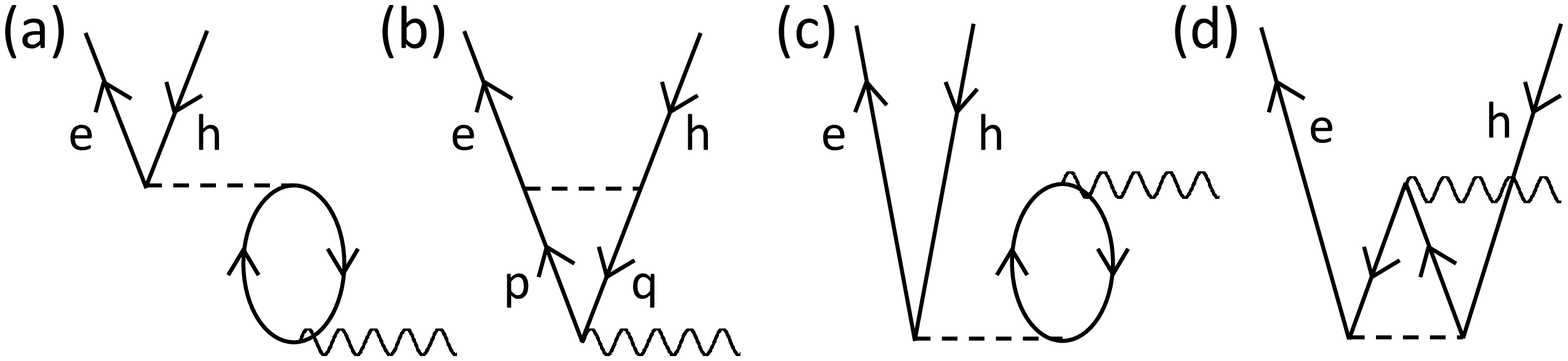}

\caption{\label{fig:first-order-mbpt}First-order Coulomb corrections to the
amplitude for one-photon interband absorption (for noninteracting
single-particle states). The final-state exciton is $(e,h)$. Upward-pointing
lines indicate states in the CB, downward-pointing lines states in
the VB.}
\end{figure}

In Fig.~\ref{fig:first-order-mbpt}, we show the the first-order
Coulomb correction to the interband absorption amplitude \cite{Mahan,LindgrenMorrison,ShavittBartlett}.
There are four time-ordered many-body diagrams in first order, as
shown in the figure. Now, in envelope-function approaches, Coulomb
matrix elements in which the band index changes at one or both vertices
are suppressed (they correspond to higher-order $\kp$ corrections),
and such diagrams vanish in the effective-mass limit (see, for example,
Ref.~\cite{Kira&Koch}). One thus expects Figs.~\ref{fig:first-order-mbpt}(a),
\ref{fig:first-order-mbpt}(c), and \ref{fig:first-order-mbpt}(d)
to be small, since the band index changes at both vertices of the
Coulomb interaction and these diagrams are therefore formally of order
$(\kp)^{2}$ or $(L_{\text{atom}}/L_{\text{dot}})^{2}$. Only Fig.~\ref{fig:first-order-mbpt}(b)
is of order $(\kp)^{0}$ and remains nonzero in the effective-mass
limit. This diagram is the \emph{vertex correction} to the electron-photon
interaction; it forms the dominant many-body correction for an interband
matrix element \cite{Mahan,elliott-57-sqd}. Note that this situation
contrasts with absorption by atoms and molecules, where all four of
the analogous diagrams give important correlation corrections \cite{LindgrenMorrison,ShavittBartlett}.
In this paper, we shall limit our discussion of correlation effects
in semiconductor NCs to Fig.~\ref{fig:first-order-mbpt}(b).

The vertex correction represents the interaction between the electron
and hole in the final state $(e,h)$ of the absorption, and it thus
accounts for the correlation in the final-state exciton wave function
in Eq.~(\ref{eq:red-amplitude}). The same physical effect in NCs
is often treated via a one-parameter variational ansatz for the exciton
wave function introduced by Takagahara \cite{TakagaharaPRB1987,BeckerNatLett2018}.
In this paper, we will instead treat the vertex correction using the
methods of MBPT, by summing over the virtual states of a HF basis
set.

The first-order vertex correction, Fig.~\ref{fig:first-order-mbpt}(b),
can be analyzed using methods of degenerate (open-shell) MBPT \cite{LindgrenMorrison,ShavittBartlett}.
The expressions for the lowest- and first-order absorption amplitude
(full, not reduced), in the case where the single-particle states
are formed in a HF potential, are given by
\begin{eqnarray}
\mathbf{M}_{eh}^{(0)} & = & \BraOperKet{e}{\mathbf{p}}{h}\,,\label{eq:M0}\\
\mathbf{M}_{eh}^{(1)}(\omega) & = & -\sideset{}{'}\sum_{pq}\frac{\BraOperKet{eq}{g_{12}}{ph}\BraOperKet{p}{\mathbf{p}}{q}}{\omega+\epsilon_{q}-\epsilon_{p}}\,.\label{eq:M1}
\end{eqnarray}
Here, $\omega$ is the excitation frequency, $p$ are CB states, $q$
are VB states, and $\epsilon_{p}$ and $\epsilon_{q}$ are the HF
eigenvalues of these states. The prime on the summation indicates
that terms are to be excluded where $p$ or $q$ is a magnetic substate
lying in the shells of the external legs ($p\notin e_{\text{shell}},$
$q\notin h_{\text{shell}}$). Note that the first-order absorption
amplitude depends on the excitation frequency $\omega$ via the energy
denominator. A similar expression for the first-order vertex correction
can be applied to emission by putting $\omega=\omega_{eh}$, the energy
of the emitted photon \footnote{The definition of the reduced amplitude given in Eq.~(\protect\ref{eq:red-amplitude})
corresponds to $\omega=\omega_{eh}$. The reduced amplitude $M_{eh}$
in Eq.~(\protect\ref{eq:cross-section}) should more correctly be
$M_{eh}(\omega)$, but the difference between using $M_{eh}(\omega)$
and $M_{eh}(\omega_{eh})$ in that equation is negligible for purposes
of the numerical applications in this paper.}.

The corresponding expression for the reduced first-order amplitude
$M_{eh}^{(1)}$ can be found by coupling the external legs of the
diagram to a total angular momentum $F_{\text{tot}}$, in analogy
with Eq.~(\ref{eq:Xeh0}). The final result is given in Appendix~\ref{app:angmom-vertex},
in the form of radial integrals and angular factors.

\begin{table}
\caption{\label{tab:example_M1}The first-order (vertex-corrected) reduced
interband momentum matrix element for the ground-state $1S_{e}$-$1S_{h}$
($F_{\text{tot}}=1$) exciton in a NC of CsPbBr$_{3}$ with edge length
$L=11$~nm, using the material parameters in Table~\protect\ref{tab:parameters}
(with $\Ep=20$~eV). Notation: $iM^{(1)}$ is the total first-order
reduced matrix element; $iM_{K}^{(1)}$ is the partial-wave contribution
to $iM^{(1)}$ arising from multipole $K$ (see Appendix~\protect\ref{app:angmom-vertex});
`extrap.' is an estimate of the extrapolated contribution from $K=13$
to infinity. Units: atomic units. The lowest-order reduced matrix
element is $iM^{(0)}=0.847$~a.u..}

\begin{ruledtabular}
\begin{tabular}{ldd}
\multicolumn{1}{l}{$K$} & \multicolumn{1}{c}{$iM_{K}^{(1)}$} & \multicolumn{1}{c}{$M_{K}^{(1)}/M^{(1)}$ (\%)}\\
\hline 
0 & 0.045 & 5.7\\
1 & 0.375 & 47.8\\
2 & 0.129 & 16.5\\
3 & 0.065 & 8.3\\
4 & 0.039 & 4.9\\
5 & 0.025 & 3.2\\
6 & 0.018 & 2.3\\
7 & 0.013 & 1.7\\
8 & 0.010 & 1.3\\
9 & 0.008 & 1.0\\
10 & 0.006 & 0.8\\
11 & 0.005 & 0.7\\
12 & 0.004 & 0.6\\
extrap. & 0.042(4) & 5.4\\
\hline 
Total $iM^{(1)}$ & 0.785(4) & \\
$i(M^{(0)}+M^{(1)})$ & 1.632(4) & \\
\end{tabular}
\end{ruledtabular}

\end{table}

An example calculation of $M_{eh}^{(1)}$ for the ground-state $1S_{e}$-$1S_{h}$
exciton in a NC of CsPbBr$_{3}$ is given in Table~\ref{tab:example_M1}.
The total angular momentum in this case can take the values $F_{\text{tot}}=1$
(bright exciton) or $F_{\text{tot}}=0$ (dark exciton), and the matrix
element shown applies to the allowed decay from $F_{\text{tot}}=1$.
The first-order matrix element is expressed as a sum over contributions
from Coulomb multipoles $K$, according to Eq.~(\ref{eq:M1-red}).
For the $1S_{e}$-$1S_{h}$ exciton, the multipole $K$ also corresponds
to the orbital angular momentum of the states $p$ and $q$ in Eq.~(\ref{eq:M1}).
For example, for $K=1$, the states $p$ and $q$ can have all combinations
of the angular momenta $P_{1/2}$ and $P_{3/2}$. The $P$-wave angular
channel can be seen to dominate the sum, accounting for about 50\%
of the matrix element. The sum over $K$ converges quite slowly, however,
with an asymptotic form approximately proportional to $1/K^{2}$;
this allows us to estimate the extrapolated contribution from $K=13$
to infinity, which is about $5\%$ of the total first-order matrix
element. In order to obtain an overall precision of better than 1\%
in the first-order matrix element, it is necessary to include the
first 9 or more principal quantum numbers in the intermediate sums
(at least, in the dominant $P$-wave channel).

The vertex correction in Table~\ref{tab:example_M1} can be seen
to be a large effect for the ground-state exciton considered here.
Including the first-order correction leads to a reduction in the radiative
lifetime by a factor of $[(M^{(0)}+M^{(1)})/M^{(0)}]^{2}\approx3.7$
relative to the HF value. These large vertex-renormalization factors
are to be expected for a NC in intermediate confinement \cite{TakagaharaPRB1987,BeckerNatLett2018},
which is the case here (see Sec.~\ref{subsec:parameters}). However,
as we shall see in Sec.~\ref{subsec:absorption-spectra}, the vertex
renormalization factors decrease rapidly as a function of excitation
energy and approach unity for excited-state excitons.

\section{\label{sec:results}Results and discussion}

\subsection{\label{subsec:parameters}Material parameters}

\begin{table}[tb]
\caption{\label{tab:parameters}Material parameters for CsPbBr$_{3}$ and CsPbI$_{3}$
used in this work. See Sec.~\protect\ref{subsec:parameters} for
explanation. $\Ep^{(4)}$ is the Kane parameter estimated from the
$4\times4$ $\mathbf{k}\cdot\mathbf{p}$ model, and $\Ep^{(8)}$ is
estimated from the $8\times8$ $\mathbf{k}\cdot\mathbf{p}$ model
{[}see Eq.~(\ref{eq:muandEp}){]}.}

\begin{ruledtabular}
\begin{tabular}{ldd}
 & \multicolumn{1}{c}{CsPbBr$_{3}$} & \multicolumn{1}{c}{CsPbI$_{3}$}\\
\hline 
$\mu^{*}$ ($m_{0}$)  & 0.126\footnotemark[1]  & 0.114\footnotemark[1]\\
$m_{e}^{*}=m_{h}^{*}$ ($m_{0}$)  & 0.252  & 0.228\\
$\Egap$ (eV)  & 2.342\footnotemark[1]  & 1.723\footnotemark[1]\\
$\Esoc$ (eV)  & 1.0\footnotemark[2]  & 1.0\footnotemark[2]\\
$\Ep^{(4)}$ (eV)  & 27.9  & 22.7\\
$\Ep^{(8)}$ (eV)  & 16.4  & 13.9 \\
$\epseff$  & 7.3\footnotemark[1]  & 10.0\footnotemark[1]\\
$\epsopt$  & 4.84\footnotemark[3]  & 4.7\footnotemark[4]\\
$\varepsilon_{\text{out}}$ & 2.4\footnotemark[5] & 2.4\footnotemark[5]\\
\end{tabular}

\footnotetext[1]{Ref.~\cite{YangACSEnergyLett2017}}

\footnotetext[2]{Ref.~\cite{YuSciRep2016}}

\footnotetext[3]{Ref.~\cite{DirinEpsVersusLambdaACSchemmater2016},
at a wavelength of 500~nm.}

\footnotetext[4]{Ref.~\cite{SinghEpsVersusLambdaJjtice2019}, at
a wavelength of 500~nm.}

\footnotetext[5]{This value is for toluene.} 
\end{ruledtabular}

\end{table}

The material parameters used in this work are summarized in Table~\ref{tab:parameters}.
We have taken the reduced mass $\mu^{*}$, the band gap $E_{g}$,
and the `effective' dielectric constant $\varepsilon_{\text{eff}}$
from Yang \emph{et al}.\ \cite{YangACSEnergyLett2017}; these were
measured at cryogenic temperatures for the orthorhombic phase of CsPbBr$_{3}$
and the cubic phase of CsPbI$_{3}$ \cite{CottinghamChemComm2016,StoumposACScg2013,HirotsuJPSJ1974}.
While $\mu^{*}=m_{e}^{*}m_{h}^{*}/(m_{e}^{*}+m_{h}^{*})$ is known,
the individual effective masses of electron $m_{e}^{*}$ and hole
$m_{h}^{*}$ are not. However, there is evidence from experiment \cite{SumTCNatCom2017}
and first-principles calculations \cite{BeckerNatLett2018,ProtesescuNanoLett2015,UmariSciRep2014}
that the effective masses are approximately equal for inorganic perovskites,
so we will assume $m_{e}^{*}=m_{h}^{*}$. The spin-orbit splitting
$\Esoc$ between the $p_{1/2}$-like and the higher-lying $p_{3/2}$-like
band is taken from Ref.~\cite{YuSciRep2016}.

The dielectric constant $\varepsilon_{\text{in}}$ used to screen
the Coulomb interactions~(\ref{eq:lrcoul}), for both the HF equations
and the vertex correction, will be taken to be the `effective' constant
$\varepsilon_{\text{in}}=\varepsilon_{\text{eff}}$ measured in Ref.~\cite{YangACSEnergyLett2017}.
The constant $\varepsilon_{\text{eff}}$ is derived from the binding
energy of the bulk exciton and therefore applies to a length scale
of order the Bohr radius $a_{B}$, which is quite close to the size
of the NCs that we calculate (using the parameters in Table~\ref{tab:parameters},
one finds $2a_{B}=6.1$~nm for CsPbBr$_{3}$ and $2a_{B}=9.3$~nm
for CsPbI$_{3}$). We also need optical dielectric constants $\varepsilon_{\text{in}}'=\varepsilon_{\text{opt}}$
to calculate the dielectric screening factor $f_{\varepsilon}$~(\ref{eq:feps-sphere}).
Note that the constant $\varepsilon_{\text{opt}}$ applies to a length
scale given by the wavelength (we take $\lambda=500$~nm, an energy
just above the threshold for absorption) and to a frequency $\omega=c/\lambda$.
Inorganic perovskites present the difficulty that the bulk dielectric
function varies rapidly with length and frequency scales, as can be
seen from the significant difference between $\varepsilon_{\text{eff}}$
and $\varepsilon_{\text{opt}}$ in Table~\ref{tab:parameters}.

Also important is the Kane parameter $\Ep$, defined by Eq.~(\ref{eq:KaneParameter}),
which serves a dual purpose in the present work: it controls the $\kp$
corrections via the coupling of the VB and the CB in the $4\times4$
$\kp$ model, and it controls the strength of the interband electron-photon
interaction, where the coupling constant is proportional to $\sqrt{\Ep}$
\cite{Mahan,elliott-57-sqd,efros-82-sqo}, as can be seen from Eq.~(\ref{eq:RME-p-diff}).
However, no direct measurements of $\Ep$ exist for CsPbBr$_{3}$
or CsPbI$_{3}$. An estimate of $\Ep$ can be made within an extended
$8\times8$ $\kp$ model, which includes the $s$-like VB ($R_{6}^{+}$)
and the $p_{1/2}$-like CB ($R_{6}^{-}$) of the $4\times4$ model,
together with the spin-orbit-split-off $p_{3/2}$-like CB band ($R_{8}^{-}$),
at the $R$ point of the Brillouin zone. If one assumes that the contribution
of remote bands to $m_{e}^{*}$ and $m_{h}^{*}$ is zero, this model
implies \footnote{This follows by putting $\gamma_{e}=1$ and $\gamma_{h}=-1$ in Ref.~\cite{EfrosRosen_annurev.matsci2000}.}
\begin{equation}
\frac{1}{\mu^{*}}=\frac{2}{3}\left(\frac{\Ep}{\Egap}+\frac{\Ep}{\Egap+\Esoc}\right)\,.\label{eq:muandEp}
\end{equation}
This equation can now be solved for $\Ep$. By allowing $\Esoc\rightarrow\infty$
in Eq.~(\ref{eq:muandEp}), one obtains the corresponding equation
\cite{even-14b-sqd,YangACSEnergyLett2017} for the $4\times4$ $\kp$
model.

The values of $\Ep$ inferred in this way for the $8\times8$ and
$4\times4$ models are summarized in Table~\ref{tab:parameters}.
We take the view that $\Ep$ is uncertain. A conservative range would
be $10\,\text{eV}\leq\Ep\leq32\,\text{eV}$ for CsPbBr$_{3}$ and
$8\,\text{eV}\leq\Ep\leq26\,\text{eV}$ for CsPbI$_{3}$. We discuss
this issue further in the next section.

\subsection{\label{subsec:results-lifetimes}Radiative lifetimes}

\begin{figure}
\includegraphics[scale=0.34]{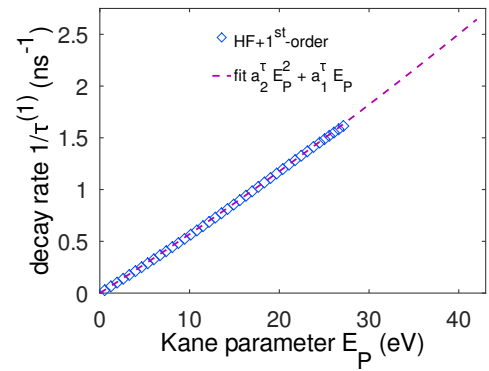}

\caption{\label{fig:decay-vs-Ep}Bright-exciton decay rate $1/\tau$ vs.\ Kane
parameter $\Ep$ for NCs of CsPbBr$_{3}$ with edge length $L=9$~nm.
Diamonds: calculated values (HF plus first-order vertex correction)
for $\Ep\alt28$~eV; dashed line: a fit to a quadratic in $\Ep$,
$1/\tau=a_{1}^{\tau}\Ep+a_{2}^{\tau}\Ep^{2}$.}
\end{figure}

The radiative decay rate of the ground-state bright exciton $1S_{e}$-$1S_{h}$
($F_{\text{tot}}=1$) in the effective-mass approximation (EMA) is
proportional to the Kane parameter $\Ep$ \cite{efros-82-sqo}. In
Fig.~\ref{fig:decay-vs-Ep}, we show the radiative decay rate of
a NC of CsPbBr$_{3}$ calculated by the present methods ($\kp$ and
MBPT), for a range of values of $\Ep$. The fit to the calculated
points is indeed quite linear, although a small curvature is present
owing to the higher-order $\kp$ corrections included in the present
approach. This approximate proportionality of the radiative decay
rate (and also of the one-photon absorption cross section) to $\Ep$
complicates quantitative comparisons between theory and experiment
while the value of $\Ep$ is uncertain. For illustrative purposes,
in subsequent figures we shall use the values $\Ep=20$~eV (CsPbBr$_{3}$)
and $\Ep=17$~eV (CsPbI$_{3}$), which are close to the average of
$\Ep^{(4)}$ and $\Ep^{(8)}$ in Table~\ref{tab:parameters}.

Note that the calculated points in Fig.~\ref{fig:decay-vs-Ep} are
for $\Ep\alt28$~eV. For higher values of $\Ep$, we find that the
$4\times4$ $\kp$ model develops unphysical intragap solutions, similar
to those encountered in $\kp$ models applied to NCs of III-V and
II-VI compounds \cite{WangPRB1999}. However, if required, it is always
possible to attempt to extrapolate physical observables to values
of $\Ep\agt28$~eV, as has been done in Fig.~\ref{fig:decay-vs-Ep}.

\begin{figure}
\includegraphics[scale=0.78]{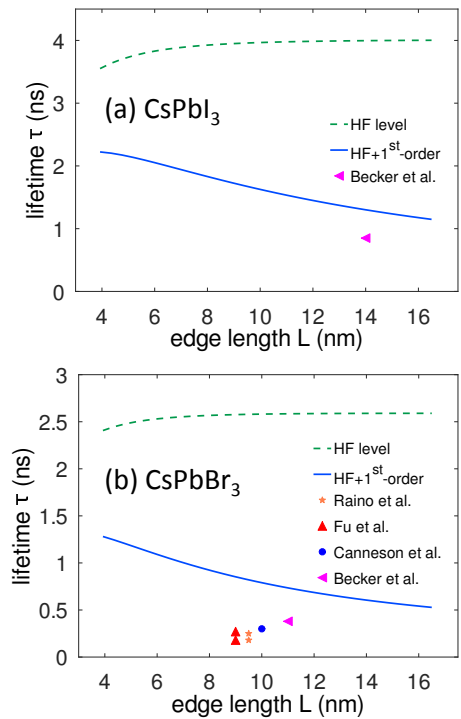}

\caption{\label{fig:lifetime-vs-L}Bright-exciton lifetime vs.\ edge length
$L$ for NCs of (a) CsPbI$_{3}$ ($\Ep=17$~eV) and (b) CsPbBr$_{3}$
($\Ep=20$~eV). Dashed curve: HF; continuous curve: HF plus first-order
vertex correction. Experimental results: Becker \emph{et al.}, Ref.~\cite{BeckerNatLett2018};
Rain{\`o} \emph{et al.,} Ref.~\cite{raino-16-sqd}; Fu \emph{et
al.}, Ref.~\cite{FuNanoLett2017}; Canneson \emph{et al.}, Ref.~\cite{CannensonNanoLett2017}.}
\end{figure}

In Fig.~\ref{fig:lifetime-vs-L}, theoretical lifetimes for NCs of
CsPbI$_{3}$ and CsPbBr$_{3}$ are shown as a function of edge length
$L$ and compared with available measurements. We note first that
the lifetime is nearly independent of $L$ in the HF approximation,
the standard result for the strong-confinement limit \cite{efros-82-sqo}.
Indeed, HF behaves like a strong-confinement theory, since the single-particle
states have the quantum numbers of the strong-confinement limit and
there is no correlation between electrons and holes (mean-field theory).
On the other hand, the vertex correction introduces correlation, which
leads to a reduction in lifetime for increasing NC size \cite{efros-82-sqo,TakagaharaPRB1987},
as can be observed in Fig.~\ref{fig:lifetime-vs-L}. However, we
find that for large NC sizes $L\gg2a_{B}$, the radiative lifetime
varies as $\tau\sim1/L$ using the present approach with a first-order
vertex correction, instead of following the dependence $\tau\sim1/L^{3}$
predicted by Refs.~\cite{efros-82-sqo,TakagaharaPRB1987}. To reproduce
this $1/L^{3}$ dependence within MBPT (using the same assumptions
as Refs.~\cite{efros-82-sqo,TakagaharaPRB1987}) evidently requires
an all-order treatment of the vertex correction. Nevertheless, the
first-order treatment used here might be expected to be a reasonable
theory in the regime of intermediate confinement, where the $L$-dependence
of the lifetime interpolates the expected $1/L^{3}$ dependence of
the weak-confinement limit and the $L^{0}$ dependence of the strong-confinement
limit. The data in Fig.~\ref{fig:lifetime-vs-L} are close to intermediate
confinement (see Sec.~\ref{subsec:parameters} for a discussion of
the Bohr radius $a_{B}$).

We note also that the measurements for CsPbBr$_{3}$ in Fig.~\ref{fig:lifetime-vs-L}
show a $\pm40$\% discrepancy among themselves. They all use toluene
as the surrounding medium, although in some cases (e.g., Ref.~\cite{BeckerNatLett2018})
there are additives. The edge length shown in the figure corresponds
to the average value of $L$ for the ensemble of NCs synthesized;
the size fluctuation is of order $\pm(0.5\text{--}1.0)$~nm for all
the measurements. The ensemble can also be expected to contain a range
of shape deformations (tetragonal, orthorhombic, and other) of the
basic cubic NC shape, and possibly different crystal phases as well
\cite{CottinghamChemComm2016,StoumposACScg2013,HirotsuJPSJ1974}.

Temperature-dependent effects can also be important. The measured
lifetimes are longer at room temperature \cite{raino-16-sqd}. At
the cryogenic temperatures used for the measurements in Fig.~\ref{fig:lifetime-vs-L},
the thermal occupation of the fine-structure states of the bright
exciton can be nonuniform. The fine-structure splittings are typically
found to be a few meV and can vary markedly from dot to dot in single-dot
measurements \cite{BeckerNatLett2018}. The lifetime may possibly
depend on the fine-structure component, depending on the origin of
the splitting, and therefore on the particular NC being investigated
(in single-dot measurements) and on the precise temperature of the
experiment; our calculation is effectively based on the assumption
of zero fine-structure splitting.

Another issue is that the measured decay rate would not be the radiative
rate if there were competing nonradiative decay channels. For instance,
if there were nonradiative decay channels directly from the bright
state, the measured lifetimes would then be too small. However, the
quantum yields are high (e.g., of order 88\% for a NC of CsPbBr$_{2}$Cl
measured in Ref.~\cite{BeckerNatLett2018}), and there is strong
evidence that the bright exciton state is the ground state in CsPbBr$_{3}$
\cite{BeckerNatLett2018}, so that nonradiative decay channels from
the bright state seem likely to be suppressed.

We see from Fig.~\ref{fig:lifetime-vs-L} that inclusion of the vertex
correction markedly improves agreement between theory and experiment.
Nevertheless, the final MBPT values of the lifetime, for the illustrative
values of the Kane parameter chosen {[}$\Ep=20$~eV (CsPbBr$_{3}$)
and $\Ep=17$~eV (CsPbBr$_{3}$){]}, still globally overestimate
the measured values, both for CsPbBr$_{3}$ and for CsPbI$_{3}$.
A simple approach would be to fit $\Ep$ to the experiments; this
would require values of $\Ep$ somewhat in excess of the value $\Ep^{(4)}$
in Table~\ref{tab:parameters} inferred from the $4\times4$ $\kp$
model. However, we note that there are other sources of theoretical
uncertainty, besides the Kane parameter. The main ones are:

(i) \emph{Uncertainty in dielectric constants}. The optical dielectric
constant $\varepsilon_{\text{in}}'=\varepsilon_{\text{opt}}$ varies
rapidly in the vicinity of the absorption threshold \cite{DirinEpsVersusLambdaACSchemmater2016,SinghEpsVersusLambdaJjtice2019}
and this influences the lifetime through the dielectric screening
factor $f_{\varepsilon}$ (\ref{eq:feps-sphere}); the values used
here (Table~\ref{tab:parameters}) correspond to a wavelength $\lambda=500$~nm.
Also, the dielectric constant of the surrounding medium would vary
if there are additives \cite{BeckerNatLett2018}; we have here assumed
the value for pure toluene. As an example of the possible effect of
these uncertainties, we note that a 15\% uncertainty in $\varepsilon_{\text{opt}}$
and a 5\% uncertainty in $\varepsilon_{\text{out}}$ would lead to
about a 17\% uncertainty in the lifetime.

In addition, the vertex-renormalization factor is sensitive to the
dielectric constant $\varepsilon_{\text{in}}$ used to screen the
Coulomb interactions (\ref{eq:lrcoul}), since the first-order Coulomb
correction is proportional to $1/\varepsilon_{\text{in}}$. We have
assumed $\varepsilon_{\text{in}}=\varepsilon_{\text{eff}}$ in our
calculations (see Sec.~\ref{subsec:parameters}). But the length
scale for the NCs is not identical to that of the bulk exciton from
which the constant $\varepsilon_{\text{eff}}$ was inferred, and the
vertex correction also samples parts of the bulk dielectric function
at nonzero frequency \cite{Mahan}, so the appropriate value of $\varepsilon_{\text{in}}$
might be somewhat different from $\varepsilon_{\text{eff}}$. As mentioned
in Sec.~\ref{subsec:parameters}, the bulk dielectric function varies
rapidly with distance and frequency. This issue is hard to quantify,
but as an example, if we assume that $\varepsilon_{\text{in}}$ is
15\% smaller than $\varepsilon_{\text{eff}}$, then the vertex renormalization
factor would increase, and the lifetime would decrease, also by about
15\%.

We note that we have also neglected the effect of the dielectric mismatch
between the NC and the surrounding medium \cite{karpulevich-19},
and that boundary effects can be further modified by the ligands \cite{karpulevich-19}.

(ii) \emph{Corrections for cubic NCs}. Although the perovskite NCs
in this study are generally cuboid, we have assumed a spherical NC
with an effective radius given by Eq.~(\ref{eq:radiusL}). As mentioned
in Sec.~\ref{subsec:model}, many errors from this approximation
are expected to enter at the few percent level for the ground-state
exciton. One source of error that we did not discuss in Sec.~\ref{subsec:model}
concerns the value of the dielectric screening factor $f_{\varepsilon}$.
The calculations above have assumed the spherical value $f_{\varepsilon}^{\text{sph}}$~(\ref{eq:feps-sphere}).
However, according to the numerical calculations for a cube in Ref.~\cite{BeckerNatLett2018},
for the case of intermediate confinement with $\varepsilon_{\text{in}}'/\varepsilon_{\text{out}}\approx2$
(as in Table~\ref{tab:parameters}), the ratio of lifetimes for a
cubic NC and a spherical NC with the same volume is $\tau^{\text{cube}}/\tau^{\text{sph}}\approx1.4$.
If instead of equal volumes we use a sphere radius given by Eq.~(\ref{eq:radiusL})
and assume that the lifetime is approximately inversely proportional
to the volume, then the ratio calculated in Ref.~\cite{BeckerNatLett2018}
is modified to $\tau^{\text{cube}}/\tau^{\text{sph}}\approx1.12$.
Thus, according to these estimates, our theoretical values for the
lifetime in Fig.~\ref{fig:lifetime-vs-L} should be increased by
about 12\%. These results also imply that $f_{\varepsilon}^{\text{sph}}/f_{\varepsilon}^{\text{cube}}\approx1.06$
for the parameters used here. Another calculation gives $f_{\varepsilon}^{\text{sph}}/f_{\varepsilon}^{\text{cube}}\approx0.99$
for $\varepsilon_{\text{in}}'/\varepsilon_{\text{out}}\approx2$ in
the strong-confinement limit \footnote{Zhe Wang, private communication (2020).}.

(iii) \emph{Higher-order MBPT}. We use a first-order vertex correction.
Unfortunately, it is difficult to estimate the effect of the omitted
higher-order vertex terms without explicit calculation. Comparing
with the results of the variational calculation in Ref.~\cite{BeckerNatLett2018},
however, we conclude that the ground-state vertex renormalization
factors (for intermediate confinement) could be increased by as much
as 40\% beyond their first-order value.

We believe that the present discrepancy between theory and experiment
is due to a combination of (iii) above and our use of the wrong value
of the Kane parameter, with further contributions from the other sources
of uncertainty (including experimental). 

\subsection{\label{subsec:absorption-spectra}One-photon absorption spectra}

\begin{figure}
\includegraphics[scale=0.62]{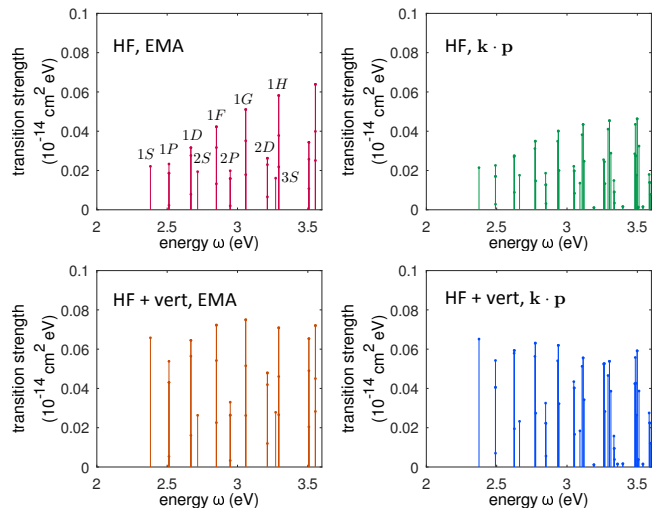}

\caption{\label{fig:transitions}Transition strengths, Eq.~(\ref{eq:transition-strength}),
in various approximations for a NC of CsPbBr$_{3}$ with edge length
$L=9$~nm ($\Ep=20$~eV). The energy axis gives the exciton energy
$\omega_{eh}=\epsilon_{e}-\epsilon_{h}$ for each possible final-state
exciton $(e,h)$. Notation: EMA, effective-mass approximation; $\kp$,
$4\times4$ $\kp$ model; HF, Hartree-Fock; HF + vert, Hartree-Fock
plus first-order vertex correction. Two levels of many-body theory
are considered: HF (upper panels) and HF with first-order vertex correction
(lower panels). For each, we employ either the EMA (left panels) or
the full $4\times4$ $\kp$ model (right panels). In the top-left
panel, the first few exciton state assignments are shown: ``$1S$''
indicates a $1S_{e}$-$1S_{h}$ exciton, ``$1P$'' a $1P_{e}$-$1P_{h}$
exciton, ``$2S$'' a $2S_{e}$-$2S_{h}$ exciton, etc. The corresponding
transitions in the other three panels have the same state assignments.}
\end{figure}

In this paper, we will only consider absorption from the highest-lying
VB ($R_{6}^{+}$) to the lowest-lying CB ($R_{6}^{-}$), around the
$R$ point of the Brillouin zone. A study of bulk excitons at cryogenic
temperatures in (CH$_{3}$NH$_{3}$)PbBr$_{3}$ \cite{tanaka-03-sqd},
which may be expected to have a band structure similar to that of
CsPbBr$_{3}$, showed a sharp excited line at 3.3~eV, which was attributed
to transitions from the $s$-like VB ($R_{6}^{+}$) to the $p_{3/2}$-like
spin-orbit-split-off CB ($R_{8}^{-}$); there were also higher-lying
structures around 3.9~eV, attributed to interband transitions at
the $M$ point. Absorption spectra of NCs of CsPbBr$_{3}$ often show
corresponding features (see, for example, Refs.~\cite{chen-17a-sqd,BrennanJACS2017}).
In particular, a step in the absorption spectrum is often visible
around 3.0--3.2~eV (for edge lengths $L\sim9$~nm), which likely
corresponds to the transition $R_{6}^{+}\rightarrow R_{8}^{-}$, in
analogy with bulk (CH$_{3}$NH$_{3}$)PbBr$_{3}$. This identification
is consistent also with density-functional (DFT) band-structure calculations
in CsPbBr$_{3}$ \cite{BeckerNatLett2018}. Because we focus on the
$R_{6}^{+}\rightarrow R_{6}^{-}$ transition here, the range of validity
of our results will extend from the absorption threshold at about
2.35~eV up to about 3.1~eV.

The first step in the calculation of the one-photon absorption cross
section (\ref{eq:cross-section}) is to calculate the reduced matrix
elements $M_{eh}$ for a large set of transitions to all possible
final-state excitons $(e,h)$ (with $F_{\text{tot}}=1$). We define
the \emph{transition strength} $T_{eh}(\omega)$ for a particular
final state $(e,h)$ to be the coefficient of the line-shape function
$\Delta_{eh}(\omega-\omega_{eh})$ in Eq.~(\ref{eq:cross-section}),
\begin{equation}
T_{eh}(\omega)=\frac{4\pi^{2}}{3}\frac{f_{\varepsilon}^{2}}{n_{\text{out}}c\omega}\left|M_{eh}\right|^{2}\,.\label{eq:transition-strength}
\end{equation}
Transition strengths $T_{eh}(\omega_{eh})$ are shown in Fig.~\ref{fig:transitions}
in various approximations. In the EMA and at HF level (top-left figure),
the dominant transitions correspond to excitons with quantum numbers
$(nl)_{e}$-$(nl)_{h}$, in which both the principal quantum number
$n$ and the orbital angular momentum $l$ of the electron and hole
are equal. Thus, the lowest-energy transition in the figure is the
$1S_{e}$-$1S_{h}$ exciton discussed in the previous section, the
next group corresponds to $1P_{e}$-$1P_{h}$ (with various $F$-dependent
fine-structure components), and so on. The selection rule on $l$
here follows directly from Eq.~(\ref{eq:RME-p-diff}); the approximate
selection rule on $n$ follows because corresponding electron and
hole wave functions are approximately equal, so that terms with $n_{e}\ne n_{h}$
are highly suppressed by the near orthogonality of the radial functions
in Eq.~(\ref{eq:RME-p-diff}).

When the HF calculation is repeated within the $4\times4$ $\kp$
model (top-right panel of Fig.~\ref{fig:transitions}), one observes
a small overall reduction in transition strength, accompanied by an
increase in the density of exciton final states. Also, the non-$S$-wave
states develop a `fine structure' corresponding to the different possible
values of total angular momentum $F$, Eq.~(\ref{eq:4compState}).
Thus, a $1P_{e}$-$1P_{h}$ exciton is split into $(1P_{1/2})_{e}$-$(1P_{1/2})_{h}$,
$(1P_{1/2})_{e}$-$(1P_{3/2})_{h}$, $(1P_{3/2})_{e}$-$(1P_{1/2})_{h}$,
and $(1P_{3/2})_{e}$-$(1P_{3/2})_{h}$ components with small energy
splittings. The fine structure is more visible for higher excited
excitons such as $1H_{e}$-$1H_{h}$. Moreover, new transitions appear
with low transition strength. This happens because the $\kp$ corrections
allow nonzero matrix elements such as $\BraOperKet{(1S_{1/2})_{e}}{\mathbf{p}}{(1D_{3/2})_{h}}$
via the `small-small' terms of Eq.~(\ref{eq:RME-p-diff}) and the
`large-small' terms of Eq.~(\ref{eq:RME-p-same}). In the EMA, this
matrix element would be forbidden because $l_{e}\ne l_{h}$.

In the lower panels of Fig.~\ref{fig:transitions}, we apply the
first-order vertex correction~(\ref{eq:M1}) to all the transitions
calculated in the upper panels \footnote{Intermediates states $(p,q)$ in Eq.~(\ref{eq:M1}) are excluded
if the energy denominator is small, $|\omega_{eq}+\epsilon_{q}-\epsilon_{p}|<\epsilon_{\text{tol}}$,
to avoid difficulties that arise in a small number of cases when two
exciton channels $(e,h)$ and $(p,q)$ have an accidental near degeneracy.
The final absorption spectrum is found to be quite insensitive to
the precise value of the cutoff $\epsilon_{\text{tol}}$ over a wide
range, e.g., $8\,\text{meV}\alt\epsilon_{\text{tol}}\alt80\,\text{meV}$.}. The vertex correction can be seen to enhance the transition strength
of the corresponding transition in the upper panel, as discussed for
the ground-state $1S_{e}$-$1S_{h}$ exciton in the previous section.
However, while the enhancement factor is large (around 3.5--4) for
the ground-state exciton, inspection of the dominant transitions in
Fig.~\ref{fig:transitions} reveals that the enhancement factor decreases
rapidly with increasing energy, so that near $\omega_{eh}=3.1$~eV,
it is much closer to unity, around 1.4, while for $\omega_{eh}=3.6$~eV,
it has decreased further to about 1.1. A simple way to understand
this result is to reflect that the Bohr radius of excited states is
larger, so that excited-state excitons are more strongly confined
than the ground-state exciton, for a given NC size.

\begin{figure}
\includegraphics[scale=0.36]{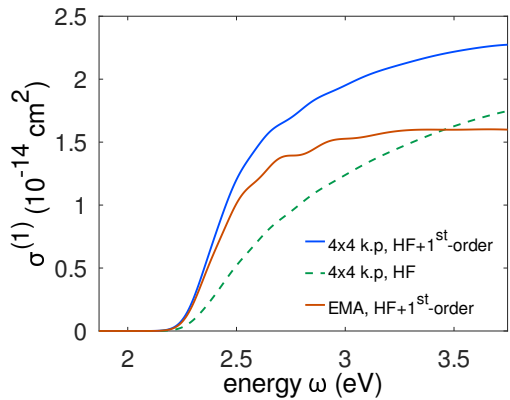}

\caption{\label{fig:sigma-MBPT-vs-HF}Calculated one-photon absorption cross
section for a NC of CsPbBr$_{3}$ with edge length $L=9$~nm ($\Ep=20$~eV)
in various approximations. Notation: HF, Hartree-Fock; EMA, effective-mass
approximation; 1st-order, first-order vertex correction.}
\end{figure}

In the next step of the calculation, we assign line-shape functions
$\Delta_{eh}(\omega-\omega_{eh})$ to each transition to produce a
broadened absorption spectrum according to Eq.~(\ref{eq:cross-section}).
In principle, the function $\Delta_{eh}(\omega-\omega_{eh})$ is a
Lorentzian for intrinsic dephasing mechanisms (homogeneous broadening),
and it is also necessary to average physical observables over the
parameters of the ensemble (inhomogeneous broadening) \cite{hu-90-sqd,hu-96-sqd}.
Here we will adopt a simpler, phenomenological approach emphasizing
inhomogeneous broadening. An important source of inhomogeneous broadening
is by the distribution of sizes in the NC ensemble. For NCs of CsPbBr$_{3}$,
the measured histogram of edge lengths $L$ can be fitted to a normal
distribution, yielding a standard deviation $\delta L$ with typical
values varying from $\delta L/L\approx5$\% \cite{chen-17a-sqd,BrennanJACS2017}
to about 10\% \cite{makarov-16-sqd,nagamine-18-sqd}. Now, since the
confinement energy is approximately proportional to $1/L^{2}$ and
the Coulomb energy to $1/L$, the exciton energy can be approximately
parametrized as
\begin{equation}
\omega_{eh}(L)=E_{g}+\frac{A_{eh}}{L^{2}}+\frac{B_{eh}}{L}\,.\label{eq:energy-L-approx}
\end{equation}
This equation, with values of $A_{eh}$ and $B_{eh}$ extracted from
the HF spectrum, may be used to relate the width $\sigma_{eh}^{\text{size}}$
of the distribution of energies $\omega_{eh}$ to the width $\delta L$
of the distribution of edge lengths $L$. The width $\sigma_{eh}^{\text{size}}$
calculated in this way is found to increase as the exciton energy
$\omega_{eh}$ increases. We then take the line-shape function to
be a Gaussian
\begin{equation}
\Delta_{eh}(\omega-\omega_{eh})=\frac{1}{\sigma_{eh}\sqrt{2\pi}}\exp\left[-\frac{(\omega-\omega_{eh})^{2}}{2\sigma_{eh}^{2}}\right]\,,\label{eq:line-shape}
\end{equation}
with $\sigma_{eh}=\sigma_{eh}^{\text{size}}$.

However, we find that size broadening alone, assuming $\delta L/L=5$--10\%,
typically produces insufficiently broadened absorption spectra containing
sharp subpeaks, which are generally not observed in measured absorption
spectra of NCs of CsPbBr$_{3}$ \cite{WangAdvMater2015,makarov-16-sqd,xu-16-sqd,chen-17a-sqd,BrennanJACS2017,nagamine-18-sqd}.
Therefore, we need to consider other broadening mechanisms. These
include phonon broadening \cite{gammon-96-sqd} and the distribution
of NC shape deformations present in the ensemble. We will treat these
effects purely phenomenologically by introducing a second width $\sigma^{\text{other}}$,
which we take to be a constant for all excitons $(e,h)$. The total
Gaussian width in Eq.~(\ref{eq:line-shape}) is then given by
\begin{equation}
\sigma_{eh}^{2}=\left(\sigma_{eh}^{\text{size}}\right)^{2}+\left(\sigma^{\text{other}}\right)^{2}\,.\label{eq:width-total}
\end{equation}
(Note that we here approximate the effect of the homogeneous phonon
broadening with a Gaussian.) A reasonable fit to the appearance of
the measured spectra can now be obtained by, for example, taking $\sigma^{\text{other}}\approx60$~meV
and $\delta L/L\approx5$\%.

One-photon absorption spectra broadened in this way are shown in various
approximations in Fig.~\ref{fig:sigma-MBPT-vs-HF}. The theoretical
spectra can be understood in terms of the underlying transition strengths
in Fig.~\ref{fig:transitions}, discussed above. After line-shape
broadening, the net effect of the $\kp$ corrections is found to be
a surprisingly large increase in the calculated cross section, reaching
about 30\% at $\omega=3.1$~eV. The cross section is also enhanced
by the vertex correction, although the enhancement factor is seen
to be much greater near the threshold than at higher energies, having
a value of only about 1.4 for the broadened cross section at $\omega=3.1$~eV.

\begin{figure}
\includegraphics[scale=0.47]{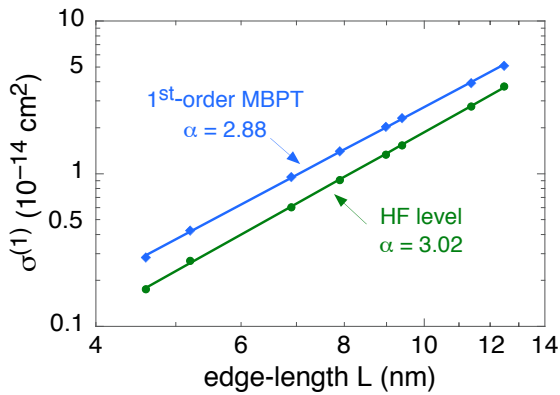}

\caption{\label{fig:sigma-vs-L}Log-log plot of the theoretical one-photon
absorption cross section at $\omega=3.1$~eV vs.\ edge length $L$
for a NC of CsPbBr$_{3}$ ($\Ep=20$~eV). Circles/Diamonds: calculated
points. Lines: fitted power-law dependence $\sigma^{(1)}(\omega)\propto L^{\alpha}$,
with the exponent $\alpha$ given by the slope of a straight-line
fit in logarithmic space.}
\end{figure}

In Fig.~\ref{fig:sigma-vs-L}, we show a log-log plot of the calculated
$\sigma^{(1)}(\omega)$ as a function of edge length $L$ at an energy
$\omega=3.1$~eV. The linearity of the log-log plot demonstrates
that the theoretical cross section for this energy fits well a power-law
dependence $\sigma^{(1)}(\omega)\propto L^{\alpha}$. A least-squares
fit to the MBPT calculation over the size range $4.5\,\text{nm}\le L\le12.5\,\text{nm}$
(Fig.~\ref{fig:sigma-vs-L}) yields a theoretical exponent $\alpha=2.88$.
This exponent agrees well with a fit to the one-photon experimental
data of Chen \emph{et al}.\ \cite{chen-17a-sqd} over the same size
range and at the same energy, which follow closely a power law with
an exponent $\alpha_{\text{expt}}=2.9\pm0.2$.

One $L$-dependent term in the theory that contributes to this exponent
is the vertex renormalization factor. We have seen, however, that
at an energy of $\omega=3.1$~eV, the vertex renormalization factor
for transition strengths is quite close to unity, of order 1.4 (Fig.~\ref{fig:transitions}),
and as a result its $L$-dependence can be expected to be a rather
weak effect. Indeed, a fit to the HF cross section (no vertex correction
present) yields a theoretical exponent $\alpha=3.02$, implying that
the vertex correction modifies the exponent by roughly $\delta\alpha_{\text{vert}}\approx-0.14$.
We conclude that the dominant $L$-dependent term is the density of
final-state excitons. In 3D, the density of states is proportional
to the volume of the confining box \cite{Mahan}, at least in the
limit of large volumes. Although the transitions (Fig.~\ref{fig:transitions})
are still quite discrete at an energy of $\omega=3.1$~eV, the line-shape
broadening discussed above yields an average density of states at
that energy. The final theoretical exponent is indeed very close to
3, particularly at HF level.

\begin{figure}
\includegraphics[scale=0.36]{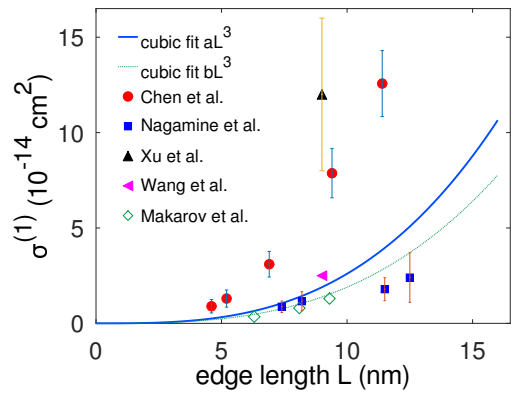}

\caption{\label{fig:sigma-vs-L-with-expts}One-photon absorption cross section
at $\omega=3.1$~eV vs.\ edge length $L$ for a NC of CsPbBr$_{3}$.
Full curves: theoretical curves ($\Ep=20$~eV) at HF level (lower
curve) and using HF plus the first-order vertex correction (upper
curve), taken from Fig.~\ref{fig:sigma-vs-L}. Circles/diamonds/triangles/squares:
experimental points. Chen \emph{et al}., Ref.~\cite{chen-17a-sqd};
Nagamine \emph{et al.}, Ref.~\cite{nagamine-18-sqd}; Xu \emph{et
al.}, Ref.~\cite{xu-16-sqd}; Wang \emph{et al.}, Ref.~\cite{WangAdvMater2015};
Makarov \emph{et al.}, Ref.~\cite{makarov-16-sqd}.}
\end{figure}

In Fig.~\ref{fig:sigma-vs-L-with-expts}, we compare our theoretical
cross section at $\omega=3.1$~eV with the results of the available
experiments that report absolute (normalized) cross sections. It can
be seen that there is significant disagreement between the various
measurements, which is probably due to uncertainties in the procedure
for normalizing the experimental cross section. The theoretical result
(for $\Ep=20$~eV) is intermediate among the various measurements.
The contributions to theoretical uncertainty discussed in Sec.~\ref{subsec:results-lifetimes}
for the lifetime apply here also, except that at an energy of $\omega=3.1$~eV,
the uncertainty due to omitted higher-order MBPT is much less, because
the vertex renormalization factors are close to unity (about 1.4),
and as a result, the calculation of the vertex factor can be expected
to be quite perturbative. Two additional error terms arise in this
case, however. First, the energy $\omega=3.1$~eV chosen for the
measurements (e.g., in Ref.~\cite{chen-17a-sqd}) is close to the
threshold for the $R_{6}^{+}\rightarrow R_{8}^{-}$ band transition,
which we have not included in our calculations. This threshold produces
a step in the cross section, which increases its value by about 20--40\%
compared to its value on the low-energy side of the step \cite{chen-17a-sqd}.

The second error term is due to the the spherical approximation. While
we pointed out in Sec.~\ref{subsec:model} that the spectra of $S$-
and $P$-like states in a cube agree well with those in the equivalent
sphere~(\ref{eq:radiusL}), the absorption cross section for $\omega\approx3.1$~eV
brings in also states of higher angular momentum, up to $G$-wave
and beyond (see Fig.~\ref{fig:transitions}). For orbital angular
momenta $l\geq2$, an $nl$ level in a sphere with degeneracy $(2l+1)$
will in general be fragmented into two or more levels in a cube, in
analogy with crystal-field theory \cite{Callaway}. Moreover, these
higher angular-momentum levels will in general be mixed by the cubic
perturbation. To estimate the overall effect of the cubic corrections,
we recalculated the absorption cross section at the level of noninteracting
particles for both a cube and a sphere, finding that $\sigma^{(1)}(\omega)$
at $\omega=3.1$~eV for a cube is greater than that for a sphere
by about 10--20\% (the precise figure being sensitive to the line-shape
function assumed).

\begin{figure}
\includegraphics[scale=0.36]{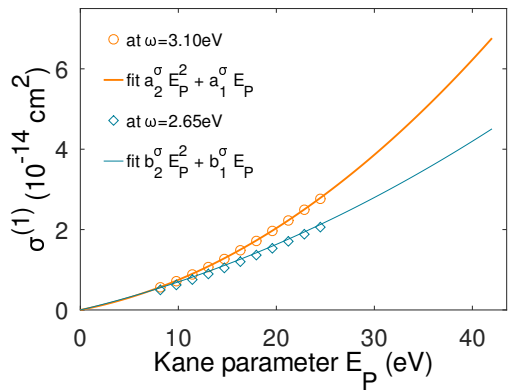}

\caption{\label{fig:sigma-vs-Ep}Calculated one-photon absorption cross section
at $\omega=3.1$~eV (circles) and $\omega=2.65$~eV (diamonds) vs.\ Kane
parameter $\Ep$ for a NC of CsPbBr$_{3}$ with edge length $L=9$~nm.
The full curves represent fits of a quadratic function of $\Ep$ through
the calculated points.}
\end{figure}

The largest theoretical uncertainty at present is, however, in the
value of the Kane parameter $\Ep$, to which the theoretical cross
section is approximately proportional (see Fig.~\ref{fig:sigma-vs-Ep}).

\section{\label{sec:conclusions} Conclusions}

We have calculated one-photon absorption cross sections for NCs of
CsPbBr$_{3}$ in various approximations, from the threshold up to
an energy of about $\omega=3.1$~eV, and compared with the available
measurements. The formalism used was a $4\times4$ $\kp$ envelope-function
model, combined with a treatment of the electron-hole correlation
within MBPT. In lowest order we used a HF model, to which we added
the first-order vertex correction to the electron-photon interaction,
which is the leading correlation correction for an interband transition.
The vertex correction gives a large enhancement, by a factor of order
3.5--4, to the absorption rate for the ground-state $1S_{e}$-$1S_{h}$
exciton, but this enhancement factor is found to decrease rapidly
as a function of excitation energy, so that for $E=3.1$~eV (about
0.7~eV above the absorption threshold), the enhancement factor is
much closer to unity, around 1.4.

The one-photon absorption cross section was obtained by computing
the transition rates to all relevant final-state excitons, with each
transition broadened phenomenologically by considering the distribution
of NC sizes in the ensemble (among other broadening mechanisms). We
gave a theoretical discussion of the absorption cross section in various
approximations, emphasizing the above-mentioned energy-dependent enhancement
by the vertex correction, as well as the effect of the $\kp$ corrections,
which turned out to be surprisingly large, yielding a 30\% enhancement
of the cross section at $E=3.1$~eV relative to a treatment within
the effective-mass approximation. The theoretical absorption cross
section at $E=3.1$~eV was shown to follow closely a power-law dependence
$\sigma^{(1)}(\omega)\propto L^{2.9}$ on the NC edge length $L$,
in close agreement with the experiment of Chen \emph{et al}.~\cite{chen-17a-sqd},
who found an exponent $\alpha_{\text{expt}}=2.9\pm0.2$. We attributed
this power-law dependence mainly to the density of final-state excitons,
with only a small contribution arising from the $L$-dependence of
the vertex-correction factors.

The available experimental data for the absolute (normalized) cross
section at $\omega=3.1$~eV show substantial disagreements among
themselves by nearly an order of magnitude; our theoretical values
(for a Kane parameter $\Ep=20$~eV) are intermediate among the measured
values. We also calculated radiative lifetimes for NCs of CsPbBr$_{3}$
and CsPbI$_{3}$, where the experimental results show a scatter of
about $\pm40$\%. Our theoretical predictions of radiative lifetimes
globally overestimate the experimental values by a factor of up to
about two (assuming $\Ep=20$~eV for CsPbBr$_{3}$ and 17~eV for
CsPbI$_{3}$).

The theoretical approach in this work can be improved in various ways.
Particularly for the radiative lifetime of the ground-state exciton,
where the first-order vertex renormalization factors are large (around
3.5--4.0), an all-order calculation of the vertex correction is clearly
indicated, even for the case of intermediate confinement encountered
in NCs of CsPbBr$_{3}$ and CsPbI$_{3}$, and should go some way toward
reducing the discrepancy with experiment observed in Fig.~\ref{fig:lifetime-vs-L}.
This involves summing numerically to all orders the electron-hole
Coulomb ladder diagrams for the final-state exciton \cite{Mahan}.
Such an all-order summation is less important, however, for the one-photon
absorption cross section at an energy $\omega=3.1$~eV, where the
corresponding vertex factors are closer to unity (around 1.4) and
the vertex correction is already quite perturbative. It would also
be interesting to improve upon the spherical NC approximation used
in this work, by adding nonspherical perturbations to the model to
take account of the cuboid NCs found for metal-halide perovskites.
Some work along these lines has been carried out in Ref.~\cite{sercel-19a-psk}.

The leading source of theoretical uncertainty at present remains the
uncertain value of the Kane parameter $\Ep$. First-principles atomistic
calculations for the bulk materials should be able to help here. A
DFT calculation for CsPbBr$_{3}$ \cite{BeckerNatLett2018} found
a reduced mass $\mu^{*}=0.065$, which is close to the result of another
DFT calculation \cite{ProtesescuNanoLett2015}, but about one half
the measured reduced mass given in Table~\ref{tab:parameters}. The
same calculation \cite{BeckerNatLett2018} found $\Ep=40$~eV, which
seems too high (at least, compared to the estimates in Table~\ref{tab:parameters}).
Further first-principles atomistic work is required to understand
the origin of these discrepancies, which may imply, for example, significant
phonon contributions to the material parameters.

\begin{acknowledgments}
The authors would like to thank Sum Tze Chien for helpful discussions.
They acknowledge the France-Singapore Merlion Project 2.05.16 for
supporting mutual visits. T.N.\ and S.B.\ are grateful to Fr{\'e}d{\'e}ric
Schuster of the CEA's PTMA program for financial support. C.G.\ gratefully
acknowledges financial support from the National Research Foundation
through the Competitive Research Program, Grant No.\ NRF-CRP14-2014-03.
\end{acknowledgments}

\appendix

\section{\label{app:momentum-mxel}Reduced momentum matrix element}

In this appendix, we derive an expression for the reduced momentum
matrix element $\RME{F_{a}}{p^{1}}{F_{b}}$ for the $4\times4$ $\kp$
model for states of spherical form~(\ref{eq:4compState}). It is
convenient for this purpose to rewrite the two-component state (\ref{eq:4compState})
in a generalized form, 
\begin{equation}
\Ket{\eta F_{a}m_{a}}=\sum_{\alpha=1}^{2}\frac{1}{r}R_{a\alpha}(r)\Ket{(l_{a\alpha},J_{\alpha})F_{a}m_{a}}\,,\label{eq:gen-2comp-state}
\end{equation}
where $\alpha=1$ or 2 denotes the component. We conventionally take
$\alpha=1$ ($\alpha=2$) to refer to the component lying in the CB
(VB). The radial function for component $\alpha$ is $R_{a\alpha}(r)$
and both components have Bloch angular momentum $J_{\alpha}=1/2$.

The matrix element $\BraOperKet{a}{\mathbf{p}}{b}$ is given by a
sum over all combinations of the component $\alpha$ of $\Ket{a}$
and the component $\beta$ of $\Ket{b}$, 
\begin{equation}
\RME{F_{a}}{p^{1}}{F_{b}}=\sum_{\alpha\beta=1}^{2}\RME{F_{a},\alpha}{p^{1}}{F_{b},\beta}\,.\label{eq:pab-sum}
\end{equation}
There are two distinct cases for $\RME{F_{a},\alpha}{p^{1}}{F_{b},\beta}$.
The first is when $\alpha\ne\beta$ (thus, $\alpha=1$ and $\beta=2$,
or $\alpha=2$ and $\beta=1$). Here, terms where $\mathbf{p}$ acts
on the envelope functions vanish, on account of the orthogonality
of the Bloch functions of the CB and VB, and therefore we only need
consider terms where $\mathbf{p}$ acts on the Bloch functions. Using
standard methods of angular-momentum theory \cite{LindgrenMorrison,Brink&Satchler},
we then find

\begin{align}
 & \RME{F_{a},\alpha}{p^{1}}{F_{b},\beta}_{\alpha\ne\beta}=\nonumber \\
 & \quad\quad(-1)^{1+F_{a}+J_{\beta}+l_{a\alpha}}\delta(l_{a\alpha},l_{b\beta})\nonumber \\
 & \quad\quad\times\sqrt{(2F_{a}+1)(2F_{b}+1)}\Sixj{F_{b}}{J_{\beta}}{l_{a\alpha}}{J_{\alpha}}{F_{a}}{1}\nonumber \\
 & \quad\quad\times\RME{J_{\alpha}}{p^{1}}{J_{\beta}}\int_{0}^{\infty}\!R_{a\alpha}(r)R_{b\beta}(r)\,dr\,.\label{eq:RME-p-diff}
\end{align}
The reduced matrix element of $p^{1}$ between Bloch states has the
value 
\begin{equation}
\RME{J_{\alpha}}{p^{1}}{J_{\beta}}=-i\sqrt{\Ep}(-1)^{L_{\alpha}}\,,\label{eq:RME-p-JaJb}
\end{equation}
where $L_{\alpha}$ is the orbital Bloch angular momentum of the band
for component $\alpha$. In lead-halide perovskites, the VB is $s$-like
and the CB is $p_{1/2}$-like, so $L_{1}=1$ and $L_{2}=0$. We define
the Kane parameter $\Ep$ by 
\begin{equation}
\Ep=2|\BraOperKet{S}{p_{z}}{Z}|^{2}\,,\label{eq:KaneParameter}
\end{equation}
where $\Ket{S}$ is the (spin-uncoupled) Bloch state of the $s$-like
band and $\Ket{Z}$ is the $z$-component of the (spin-uncoupled)
Bloch state of the $p$-like band \footnote{The Kane parameter $\Ep$ is sometimes defined to be 1/3 of this value
in the context of the $4\times4$ $\kp$ model. See, for example,
Ref.~\cite{YangACSEnergyLett2017}.}.

The second case arising in Eq.~(\ref{eq:pab-sum}) is when $\alpha=\beta$
(thus, $\alpha=\beta=1$ or $\alpha=\beta=2$). In this case, terms
where $\mathbf{p}$ acts on Bloch functions vanish, because we are
assuming the bands to have exact inversion symmetry. Therefore, we
only need consider terms where $\mathbf{p}$ acts on the envelope
functions. This gives 
\begin{align}
 & \RME{F_{a},\alpha}{p^{1}}{F_{b},\beta}_{\alpha=\beta}=\nonumber \\
 & \quad\quad\quad(-1)^{1+F_{b}+J_{\alpha}+l_{a\alpha}}\left(\frac{1}{m_{\alpha}^{*}}\right)'\nonumber \\
 & \quad\quad\quad\times\sqrt{(2F_{a}+1)(2F_{b}+1)}\Sixj{F_{b}}{l_{b\beta}}{J_{\alpha}}{l_{a\alpha}}{F_{a}}{1}\nonumber \\
 & \quad\quad\quad\times(-i)\RME{l_{a\alpha}}{\nabla^{1}}{l_{b\beta}}\,,\label{eq:RME-p-same}
\end{align}
where the reduced matrix element of the gradient operator between
envelope functions is given by 
\begin{align}
 & \RME{l_{a\alpha}}{\nabla^{1}}{l_{b\beta}}=\sqrt{l_{b\beta}+1}\nonumber \\
 & \quad\quad\quad\times\int_{0}^{\infty}\!R_{a\alpha}(r)\left[\frac{dR_{b\beta}(r)}{dr}-(l_{b\beta}+1)\frac{R_{b\beta}(r)}{r}\right]\,dr\label{eq:RME-grad+lbp1}
\end{align}
when $l_{a\alpha}=l_{b\beta}+1$, and by 
\begin{align}
 & \RME{l_{a\alpha}}{\nabla^{1}}{l_{b\beta}}=-\sqrt{l_{b\beta}}\nonumber \\
 & \quad\quad\quad\times\int_{0}^{\infty}\!R_{a\alpha}(r)\left[\frac{dR_{b\beta}(r)}{dr}+l_{b\beta}\frac{R_{b\beta}(r)}{r}\right]\,dr\label{eq:RME-grad+lbm1}
\end{align}
when $l_{a\alpha}=l_{b\beta}-1$, and $\RME{l_{a\alpha}}{\nabla^{1}}{l_{b\beta}}$
is zero in all other cases.

Equation~(\ref{eq:RME-p-same}) contains an extra complication, the
reduced-mass factor. It is well known (see, for example, Ref.~\cite{Kira&Koch})
that in an effective-mass model, with the VB and CB uncoupled, the
inclusion of $\kp$ corrections leads to an extra factor of $(1/m_{\alpha}^{*})$
multiplying the intraband momentum matrix element, where $m_{\alpha}^{*}$
is the band effective mass. This factor is in general large for a
semiconductor (e.g., it has the value $1/m_{\alpha}^{*}\approx4$
for CsPbBr$_{3}$ and CsPbI$_{3}$) and can not normally be neglected.
An analogous argument applies to the case $\alpha=\beta$ above, except
that in our coupled VB-CB model, the contributions to the effective
masses arising from the VB-CB coupling are included automatically
in the formalism. Therefore, we require instead a modified factor
$(1/m_{\alpha}^{*})'$ that includes only the contributions of the
remote bands and the bare electron mass. In the $4\times4$ $\kp$
model, this modified factor for the CB ($\alpha=1)$ is given by \cite{EfrosRosen_annurev.matsci2000}
\begin{equation}
\left(\frac{1}{m_{1}^{*}}\right)'=\frac{1}{m_{e}^{*}}-\frac{\Ep}{3E_{g}}\,,\label{eq:mfac-1}
\end{equation}
while for the VB ($\alpha=2$) 
\begin{equation}
\left(\frac{1}{m_{2}^{*}}\right)'=-\left(\frac{1}{m_{h}^{*}}-\frac{\Ep}{3E_{g}}\right)\,,\label{eq:mfac-2}
\end{equation}
where $m_{e}^{*}$ and $m_{h}^{*}$ are the full electron and hole
effective masses, respectively (which are conventionally defined to
be positive). Note that we have included an overall minus sign in
the definition of $(1/m_{2}^{*})'$ for the VB in Eq.~(\ref{eq:mfac-2});
this is needed because the matrix element in Eq.~(\ref{eq:RME-p-same})
is defined to apply to \emph{electron} states $a$ and $b$, even
when the states lie in the VB. It is now possible to show, analytically
or numerically, that the definitions (\ref{eq:mfac-1}) and (\ref{eq:mfac-2}),
together with the matrix elements (\ref{eq:RME-p-diff}) and (\ref{eq:RME-p-same}),
imply that in the uncoupled model one recovers the standard factor
$(1/m_{\alpha}^{*})$ for the intraband momentum matrix element.

Equations~(\ref{eq:pab-sum}), (\ref{eq:RME-p-diff}), and (\ref{eq:RME-p-same})
define the complete reduced matrix element. In this paper we only
need interband matrix elements, but we emphasize that the same equations
apply to both interband and intraband matrix elements, although different
terms dominate in each case. For instance, consider an interband matrix
element $\BraOperKet{a}{\mathbf{p}}{b}$, where $a$ is a CB state
and $b$ is a VB state. Then, component 1 of $\Ket{a}$ is the large
component, and component 2 is the small component; for $\Ket{b}$,
the large and small components are reversed. It follows that the large-large
($\alpha=1$, $\beta=2$) term of Eq.~(\ref{eq:RME-p-diff}) is the
dominant term, while the small-small ($\alpha=2$, $\beta=1$) term
is an $O(\kp)^{2}$ correction, and all the large-small terms of Eq.~(\ref{eq:RME-p-same})
are $O(\kp)$ corrections.

On the other hand, consider an intraband matrix element $\BraOperKet{a}{\mathbf{p}}{b}$,
where both $a$ and $b$ are states in the CB. Now the dominant term
is the large-large ($\alpha=1$, $\beta=1$) term from Eq.~(\ref{eq:RME-p-same}),
while the small-small ($\alpha=2$, $\beta=2$) terms are $O(\kp)^{2}$
corrections, and all the large-small terms from Eq.~(\ref{eq:RME-p-diff})
are $O(\kp)$ corrections.

\section{\label{app:angmom-vertex}Angular reduction of vertex correction}

To perform the angular reduction of the vertex correction, we couple
the final-state exciton $(e,h)$ in Eq.~(\ref{eq:M1}) to a total
angular momentum $F_{\text{tot}}$, as was done for the lowest-order
amplitude in Eq.~(\ref{eq:Xeh0}), and then perform the sums over
the magnetic substates analytically \cite{LindgrenMorrison,Brink&Satchler}.
This gives
\begin{eqnarray}
M_{eh}^{(1)}(\omega) & = & \delta(F_{\text{tot}},1)\sideset{}{'}\sum_{pq}\sum_{K=0}^{\infty}(-1)^{F_{p}+F_{q}}\Sixj{K}{F_{p}}{F_{e}}{1}{F_{h}}{F_{q}}\nonumber \\
 &  & \quad\quad\quad\quad\quad\quad\times\frac{X_{K}(eqph)\RME{p}{p^{1}}{q}}{\omega+\epsilon_{q}-\epsilon_{p}}\,,\label{eq:M1-red}
\end{eqnarray}
where $\RME{p}{p^{1}}{q}$ is the reduced single-particle momentum
matrix element discussed in Appendix~\ref{app:momentum-mxel}, and
$X_{K}(eqph)$ is a reduced two-body Coulomb matrix element with multipole
$K$, which is defined by 
\begin{multline}
\BraOperKet{ab}{g_{12}}{cd}=\sum_{K=0}^{\infty}\sum_{M=-K}^{K}(-1)^{F_{a}+F_{b}+K-m_{a}-m_{b}-M}\\
\shoveleft\quad\times\Threej{F_{a}}{K}{F_{c}}{-m_{a}}{M}{m_{c}}\Threej{F_{b}}{K}{F_{d}}{-m_{b}}{-M}{m_{d}}X_{K}(abcd)\,.\label{eq:red2body}
\end{multline}
In the approximation that one neglects the small components of the
states, the expression for $X_{K}(abcd)$ is analogous to the standard
expression for an atom \cite{LindgrenMorrison}, 
\begin{align}
 & X_{K}(abcd)=\frac{(-1)^{K}}{\varepsilon_{\text{in}}}\RME{\kappa_{a}}{C^{K}}{\kappa_{c}}\RME{\kappa_{b}}{C^{K}}{\kappa_{d}}\nonumber \\
 & \quad\quad\times\int_{0}^{\infty}\!\int_{0}^{\infty}\!\left(R_{a}R_{c}\right)_{r_{1}}\left(R_{b}R_{d}\right)_{r_{2}}\frac{r_{<}^{K}}{r_{>}^{K+1}}\,dr_{1}dr_{2}\,.\label{eq:Xk}
\end{align}
Here $R_{m}(r)$ is the radial function of the large component of
state $m$, and $\RME{\kappa_{a}}{C^{K}}{\kappa_{c}}$ is a reduced
matrix element of the $C^{K}$ tensor between coupled spinors \cite{Edmonds}
\begin{equation}
\RME{\kappa_{a}}{C^{K}}{\kappa_{c}}=\RME{(l_{a},1/2)F_{a}}{C^{K}}{(l_{c},1/2)F_{c}}\,.\label{eq:Ck-red}
\end{equation}
$X_{K}(abcd)$ can also be generalized to include both large and small
components by exploiting the analogy between the $4\times4$ $\kp$
model and the Dirac equation and using the techniques described in,
for example, Ref.~\cite{johnson-88a-mbpt}. We have included the
small components in the numerical calculations in this paper.

In practice, the allowed multipoles of $X_{K}(abcd)$ are limited
by parity and angular-momentum selection rules.


\bibliography{one-photon_purged}

\end{document}